\documentclass[twocolumn,showpacs,amsmath,amssymb]{revtex4-1}

\usepackage{graphicx}
\usepackage{dcolumn}
\usepackage{bm,color}

\newcommand{\Eref}[1]{Eq.~(\ref{#1})}
\newcommand{\Fref}[1]{Fig.~\ref{#1}}
\newcommand{\figwidth}{3.35in}

\begin{document}

\preprint{APS/123-QED}

\title{Partial annealing of a coupled mean-field spin-glass model with
an embedded pattern}

\author{Ayaka Sakata}
\email{ayaka@huku.c.u-tokyo.ac.jp}
\affiliation{Department of Basic Science, Graduate school of Arts and Sciences, The University of Tokyo, Komaba, Meguro-ku, Tokyo 153-8902, Japan.}
\author{Koji Hukushima}%
\email{hukusima@phys.c.u-tokyo.ac.jp}
\affiliation{Department of Basic Science, Graduate school of Arts and Sciences, The University of Tokyo, Komaba, Meguro-ku, Tokyo 153-8902, Japan.}

\date{\today}

\begin{abstract}
A partially annealed mean-field spin-glass model
with a locally embedded pattern is studied.
The model consists of two dynamical variables, spins and interactions,
 that are in contact with thermal baths at temperatures $T_S$ and $T_J$, respectively.
Unlike the quenched system, characteristic correlations among the
 interactions are induced by the partial annealing. 
The model exhibits three phases, which are paramagnetic, ferromagnetic
 and spin-glass phases. In the ferromagnetic phase, the embedded pattern
 is stably realized. 
The phase diagram depends significantly  on the ratio of two
 temperatures $n=T_J/T_S$. 
In particular, a reentrant transition from the embedded ferromagnetic
to the spin-glass phases with $T_S$ decreasing is found only below 
at a certain value of $n$. This indicates that above the critical value
 $n_c$ the embedded pattern is supported by local field from
a non-embedded region. 
Some equilibrium properties of the interactions in the partial annealing
 are also discussed in terms of frustration. 
\end{abstract}

\maketitle

\section{Introduction}

Spin-glass (SG) models \cite{beyond,Nishimori} have been extensively applied
to widespread fields for describing random or randomized spin systems. 
In such a system, there are two dynamical variables,  fast and slow
variables that are spins and interactions in a spin model, respectively.    
The latter is usually assumed to be quenched for simplicity, and is 
distributed independently and identically according
to a given distribution. 
This is based on the fact that the relaxation time of the interactions
is considerably larger than that of the spins. 
As a result of the quenched randomness, the state of the interactions is 
not affected at all by the fluctuation of spins. 
The randomness of the interactions causes many competitions between the spins,
and sometimes there is no way for eliminating them completely.
This competition that is called frustration provides rich and 
interesting phenomena in the random systems.

%

The quenched randomness is an appropriate assumption for a spin-glass
problem as a magnetic material.  
In some interesting systems, however, 
the fluctuation of the spin variables has large influence on the
interaction variables.
For example, the interactions of amino-acid sequences 
provide a globally stable state in the protein. 
The existence of such global attraction in the folding process was proposed as a consistency principle \cite{Go} or funnel
landscape \cite{Onuchic}. 
The characteristics of interactions are not expected in randomly
constructed interactions.
It is supposed that such characteristics are acquired through
an evolutional process under some fluctuations of the fast variables.
In the case of the protein, 
the fluctuation is due to the folding dynamics of the amino-acid sequences.
Similar properties have also been discovered in gene regulatory networks \cite{FangLi-Ouyang-Tang}
and transcriptional networks \cite{Alon}. 
The formation mechanism of the funnel landscape is still not fully
understood.    

Recently, an adiabatic two-temperature spin model has
been studied as a model of evolution by using Monte Carlo
simulations \cite{SHK,SHK-PRE}. 
In the model, interactions evolved so as to increase the probability
of spins to find a specific pattern of local spins.
Interestingly,  adapted interactions that exhibit funnel-like dynamics with some
robustness have been found only in an intermediate temperature
region. 
This type of interactions has no frustration around the local
spins and retains some frustrations in the rest.
The fact implies that the local
spin pattern is stabilized in the adapted interactions by energetic
and entropic effects. 
Note that there is no explicit driving force for
constructing interactions outside the local spins. 
Unfortunately, since numerical simulations are often hampered
by finite-size effects,  thermodynamic properties and phase structure
have not yet been understood well. We consider it worthwhile to clarify
the nature of such self-organized interactions with an local spin pattern
under the thermal fluctuations from the viewpoint of a
statistical-mechanics. 

In contrast to the quenched system, a feedback from the fast variables
to the slow ones is taken into account for studying such systems. 
The feedback effect is essential for the emergence of some functional
feature in the evolutional process. 
Under the assumption of the complete separation of time-scales between
the fast and slow variables, the adiabatic elimination of fast variables has been employed 
in non-equilibrium and non-linear physics \cite{MoriZwanzig,KanekoPTP}.
One of such approaches,  called partial annealing \cite{Dotsenko},  
is  introduced by Penney et. al. \cite{PCS}, in which the fast and slow 
variables touch with different heat baths and a non-equilibrium system
is mapped onto a particular equilibrium statistical-mechanical system
with two temperatures.    
The partial annealing approach has been applied to many topics; a model
of glasses \cite{Saakian}, charged system \cite{Mamasakhlisov},
spin-lattice gas \cite{CHNakajima}, 
neural network \cite{Uezu}, protein-folding \cite{Rabello}, liquid
crystal \cite{LiquidCrystal} and evolution \cite{Wagner,Ancel-Fontana}.


In this paper, we study a coupled mean-field
spin-glass model \cite{Takayama-model} in the partial annealing.  
In this model, 
two fully-connected mean-field systems are coupled with 
each other through spin-glass interactions.
One of them is a
ferromagnetic system regarded as a local region in which a
ferromagnetic pattern is embedded. 
The other is a spin-glass system providing redundant 
interactions together with the coupled interactions. 
We formulate the mean-field theory for this model in partial annealing
and discuss how the redundant interactions are modified by the thermal
fluctuation of spins. 

This paper is organized as follows.
In section \ref{model-def}, we introduce a mean-field spin-glass model
studied in detail and give a
theoretical formulation of the model in the partial annealing by using the
replica method. 
In section \ref{sec:results}, we present phase diagram of the model and discuss the equilibrium state of interactions through frustration.
Finally, section \ref{conclusion} is devoted to conclusions and outlook for further developments.

\section{Model and replica method}
\label{model-def}

\begin{figure}
\begin{center}
\includegraphics[width=\figwidth]{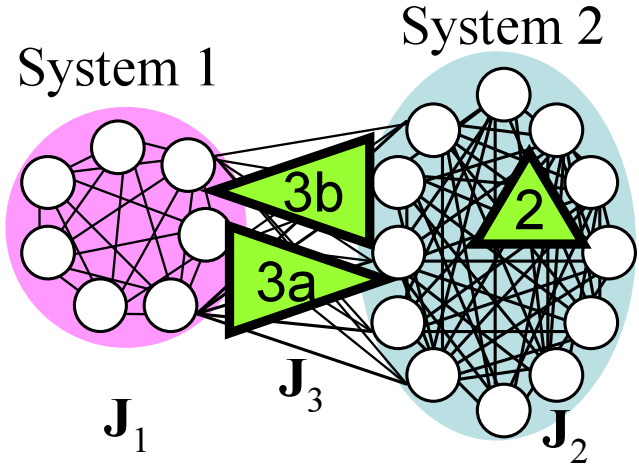}
\end{center}
\caption{(Color online) A schematic representation of a coupled
 mean-field spin-glass model. Spins and interactions are defined on
 vertexes denoted by circles and edges between vertexes, respectively. 
Three triangles represent index of frustration parameters, discussed in Sec.~\ref{simulation}.}
\label{takayama-frust}
\end{figure}

We study a coupled mean-field spin-glass model of two fully connected
spin systems,  which consists of $N_1$ Ising spins $\{S_{1,i}\}(i=1,\cdots,
N_1)$ and $N_2$ Ising spins $\{S_{2,i}\}(i=1,\cdots,N_2)$. 
The spin variables of the two systems 
and the interactions between
two spins are denoted by $\bm{S}$ and $\bm{J}$ for short,
respectively. 
The Hamiltonian of the model is given by
\begin{align}
H(\bm{S}|\bm{J})=-\sum_{p=1}^2\sum_{i<j}J_{p,ij}S_{p,i}S_{p,j}-\sum_{i,j}J_{3,ij}S_{1,i}S_{2,j}, 
\end{align}
where the summation in the first term is over all pairs within each
spin system and the summation in the second term is over all $i$ and
$j$. 
$\{J_{1,ij}\}$ and $\{J_{2,ij}\}$, denoted hereafter by $\bm{J}_1$ and
$\bm{J}_2$, respectively, are the intra-interactions within each system
and $\{J_{3,ij}\}$ denoted by $\bm{J}_3$ are inter-interactions between
these two systems. 
A schematic picture of the model is shown in Fig.~\ref{takayama-frust}. 
In a quenched system, 
they are assumed to be independently and identically distributed
according to the Gaussian distribution with mean $J_{0,p}$ and variance
$J_p$,
\begin{align}
P_0(J_{p,ij})=\sqrt{\frac{N_p}{2\pi
 (J_p)^2}}\exp\Big\{-\frac{N_p}{2(J_p)^2}\Big(J_{p,ij}-\frac{J_{0,p}}{N_p}\Big)^2\Big\}, 
\label{eqn:bare}
\end{align}
where the order of interactions $J_{p,ij}$ is scaled as $N_p^{1/2}$ for
keeping 
the Hamiltonian extensive
and $N_3$  is set to be the
geometrical mean of $N_1$ and $N_2$ as $N_3=\sqrt{N_1N_2}$ \cite{Takayama-model}. 

In this paper, we consider that an ordering pattern is embedded in one
of the system (system 1) and no a prior pattern is introduced in the
other system (system 2). This is represented  by a specific case of the
model Hamiltonian, in which the system 1 and the system 2 are a pure
ferromagnetic and spin-glass systems, respectively, and they are coupled
with each other through spin-glass interactions. Explicitly, the model
is given as 
\begin{align}
\nonumber
&J_{0,1}=J_0~(>0),~ J_{0,2}=J_{0,3}=0,\\
&J_{1}=0,~ J_2=J_3=J~(>0). 
\end{align}
Note that in the model with these parameters the interactions $\bm{J}_2$
and $\bm{J}_3$ are expected to be modified by the partial annealing
while $\bm{J}_1$ is kept to be fixed to the pure ferromagnetic interaction. 
A size ratio $r$ between the two systems is defined as 
\begin{align}
N_1/N_2=1/r^2,
\label{r}
\end{align} 
yielding that $N_1 = N\slash(1+r^2)$ and $N_2=Nr^2\slash(1+r^2)$  with
the total number of spins $N=N_1+N_2$. 
Two limiting cases $r=0$ and $r=\infty$ correspond to the
Husimi-Temperley  model and the Sherrington-Kirkpatrick model,
respectively. 


In the partial annealing system, the interactions $\bm{J}$ as well as
the spin variables $\bm{S}$ are treated as a dynamical variable. Time
scales associated with $\bm{J}$ is extremely slow and it is
assumed that the time scale is separated from that of the spin
variables. Then, the equilibrium distribution of the spins at an inverse
temperature $\beta_S=1/T_S$ is given by 
\begin{equation}
P(\bm{S}|\bm{J}) =
 \frac{1}{Z(\bm{J})}\exp(-\beta_SH(\bm{S}|\bm{J}))\label{S_equil} , 
\end{equation}
where $Z(\bm{J})$ is a partition function under a given $\bm{J}$.  
The distribution function of $\bm{J}$ at an inverse temperature
$\beta_J=1/T_J$, different from $\beta_S$, is given by
\begin{equation}
 P(\bm{J}) = \frac{1}{\cal Z}\exp(-\beta_JH_J), 
\label{eqn:eqJ}
\end{equation}
where $H_J$ is a Hamiltonian of $\bm{J}$ and ${\cal Z}$ is the total
partition function. The Hamiltonian of $\bm{J}$ is generally expressed in terms of equilibrium quantities of $\bm{S}$ and the bare distribution
$P_0(\bm{J})$ of \Eref{eqn:bare}. Although the explicit form of $H_J$
can be arbitrary chosen,  
in this study, as in \cite{PCS,Dotsenko}, we set it as 
\begin{equation}
H_J(\bm{J})=F(\bm{J})-T_J\log P_0(\bm{J}),
\label{HJ_def}
\end{equation}
where $F(\bm{J})$ is the free energy defined by $F(\bm{J})=-T_S\log Z(\bm{J})$ and
$P_0(\bm{J})=\prod_{i<j}P_0(J_{ij})$. 

Then, the equilibrium distribution $P(\bm{J})$ and the total partition
function are rewritten as
\begin{eqnarray}
 P(\bm{J}) & = & \frac{1}{\cal Z} P_0(\bm{J}) Z^n(\bm{J}),\\
{\cal Z} & = & \left[Z^n(\bm{J})\right]_0, 
\end{eqnarray}
where $n$ is the ratio between two temperatures, $n=T_S/T_J$, and
$[\cdots]_0$ means the average over $\bm{J}$ according to the bare 
distribution $P_0(\bm{J})$. 
When $n=0$, the distribution $P(\bm{J})$ is identical to $P_0(\bm{J})$
and the system corresponds to the quenched one.  For finite $n$ and
$\beta_J$, the interactions $\bm{J}$ with a lower free energy likely
occur. 

In the quench limit, the model is reduced to that studied by
Takayama \cite{Takayama-model0,Takayama-model}.  
The coupled mean-field model has been introduced in order to study
inhomogeneity of interactions between spins in real SG
materials \cite{Takayama-model0,Takayama-model}. 
Thus, the previous studies focused attention only on the
quenched system. In this work, we rather pay attention to
the system with the partial annealing, in which the interactions except for the
embedded one in system 1 are adiabatically affected by fluctuations of
spins. Our main purpose is to study the partial annealing effect on the
stability of the embedded ferromagnetic ordering in the system 1 and to
clarify characteristics of the resultant interactions by the partial
annealing.

The total free energy per spin $f$ at two inverse temperatures
$\beta_S$ 
and $\beta_J$ 
can be written as 
\begin{equation}
  f(T_S,T_J) = -\frac{1}{\beta_S} \lim_{N\rightarrow\infty}
   \frac{1}{N}\log\left[Z^n(\bm{J})\right]_0.  
\end{equation}
Following the standard procedure of the replica method \cite{beyond}, the quantity
$[Z^n]_0$ is calculated for a positive integer $n$ and an analytic
continuation to a real value given by two temperatures is taken after the calculation. Within the
assumption of replica symmetry, the free-energy density is described in
terms of order parameters $q_1$, $q_2$ and $m_1$,  and their conjugate 
parameters $\hat{q}_1$, $\hat{q}_2$ and $\hat{m}_1$, as  
\begin{align}
\nonumber
f(T_S,T_J)=-T_S\Big[-\frac{n(n-1)}{2}\sum_{p=1}^2n_p q_p \hat{q}_p
 -n_1nm_1\hat{m}_1 \\
\nonumber
+\frac{J_0\beta_Sn_1n}{2}m_1^2+\frac{J^2\beta_S^2n_2n(n-1)}{4}q_2^2+\frac{J^2\beta_S^2n_2}{4}\\
\nonumber
+\frac{\hat{n}_3J^2\beta_S^2n(n-1)}{2}q_1q_2+\frac{J^2\beta_S^2n\hat{n}_3}{2}\\
+n_1\log\int
 Dz~2\cosh^n(\sqrt{\hat{q}_1}z+\hat{m}_1)-\frac{\hat{q}_1nn_1}{2}\nonumber \\
+n_2\log\int
 Dz~2\cosh^n(\sqrt{\hat{q}_2}z)-\frac{\hat{q}_2nn_2}{2}\Big]
\label{FreeEnergy}
\end{align}
where $n_p=N_p\slash N$ $(p=1,2)$,
$\hat{n}_3=
N_3\slash N$ 
and $\int Dz=\int_{-\infty}^{+\infty} dz~e^{-z^2\slash 2}\slash\sqrt{2\pi}$.
The order parameters follow the self-consistent equations, 
\begin{align}
q_p(T_S,n)=\frac{\int Dz\tanh^2\Xi_p\cosh^n\Xi_p}{\int Dz\cosh^n\Xi_p}
\label{saddle_q}
\end{align}
and
\begin{align}
m_1(T_S,n)=\frac{\int Dz\tanh\Xi_1\cosh^n\Xi_1}{\int Dz\cosh^n\Xi_1},
\label{saddle_m}
\end{align}
where $\Xi_1=\sqrt{\hat{q}_1}z+\hat{m}_1$ and $\Xi_2=\sqrt{\hat{q}_2}z$. 
The conjugate parameters, $\hat{q}_1$, $\hat{q}_2$ and $\hat{m}_1$, are then given by
\begin{align}
\hat{q}_1=\beta_S^2J^2(rq_2),~\hat{q}_2=\beta_S^2J^2(q_1\slash
 r+q_2)\label{q_hat}, 
\hat{m}_1=\beta_SJ_0m_1.  
\end{align}

By solving the self-consistent equations we have the following solutions: 
paramagnetic solution $(q_1=q_2=m_1=0)$, ferromagnetic one $(q_1>0,q_2>0,m_1>0)$, and spin-glass one $(q_1>0,q_2>0,m_1=0)$. 
The transition temperatures between two of the phases corresponding to
these solutions are given by 
\begin{align}
T_S^{\mbox{\tiny FM-SG}}&=J_0(1+(n-1)q_1), \label{FM-SG}\\
T_S^{\mbox{\tiny PM-SG}}&=\sqrt{(1+\sqrt{5})\slash2}J\equiv\sqrt{\alpha} J, \\
T_S^{\mbox{\tiny PM-FM}}&=J_0, 
\end{align}
where PM, FM and SG mean paramagnetic, ferromagnetic, and spin-glass
phase, respectively. 
The phase boundary between the paramagnetic phase and the spin-glass one
and that between the paramagnetic phase and the ferromagnetic one are
independent of $r$ and $n$.  
Thus, the multicritical point located on $(J_0\slash J,~T_S\slash J)=(\sqrt{\alpha},\sqrt{\alpha})$ in
the $J_0\slash J-T_S\slash J$ plane is independent of $r$ and $n$.
They are consistent with the previous work in the
quench limit \cite{Takayama-model}. 

By following the stability analysis of the RS solutions by de Almeida
and Thouless (AT) \cite{AT}, the stability condition is derived as 
\begin{align}
\nonumber
1-\beta_S^2J^2&(1-2q_2+r_2)\\
&-\beta_S^4J^4(1-2q_1+r_1)(1-2q_2+r_2)>0,
\label{AT}
\end{align}
where 
\begin{align}
r_p=\frac{\int Dz\tanh^4\Xi_p\cosh^n\Xi_p}{\int Dz\cosh^n\Xi_p},~(p=1,2).
\end{align}


\section{Results}
\label{sec:results}
\subsection{Phase diagram}
\label{sec:reentrant}
\begin{figure}
\begin{center}
\includegraphics[width=\figwidth]{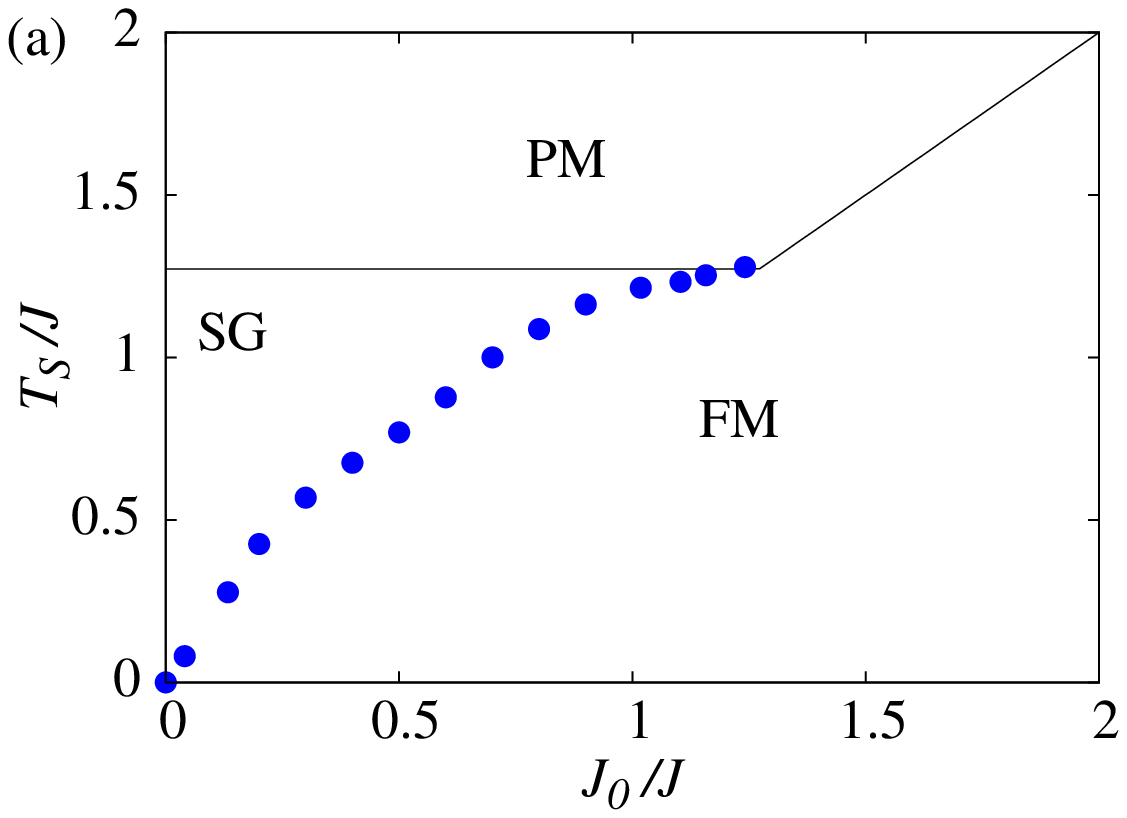}
\includegraphics[width=\figwidth]{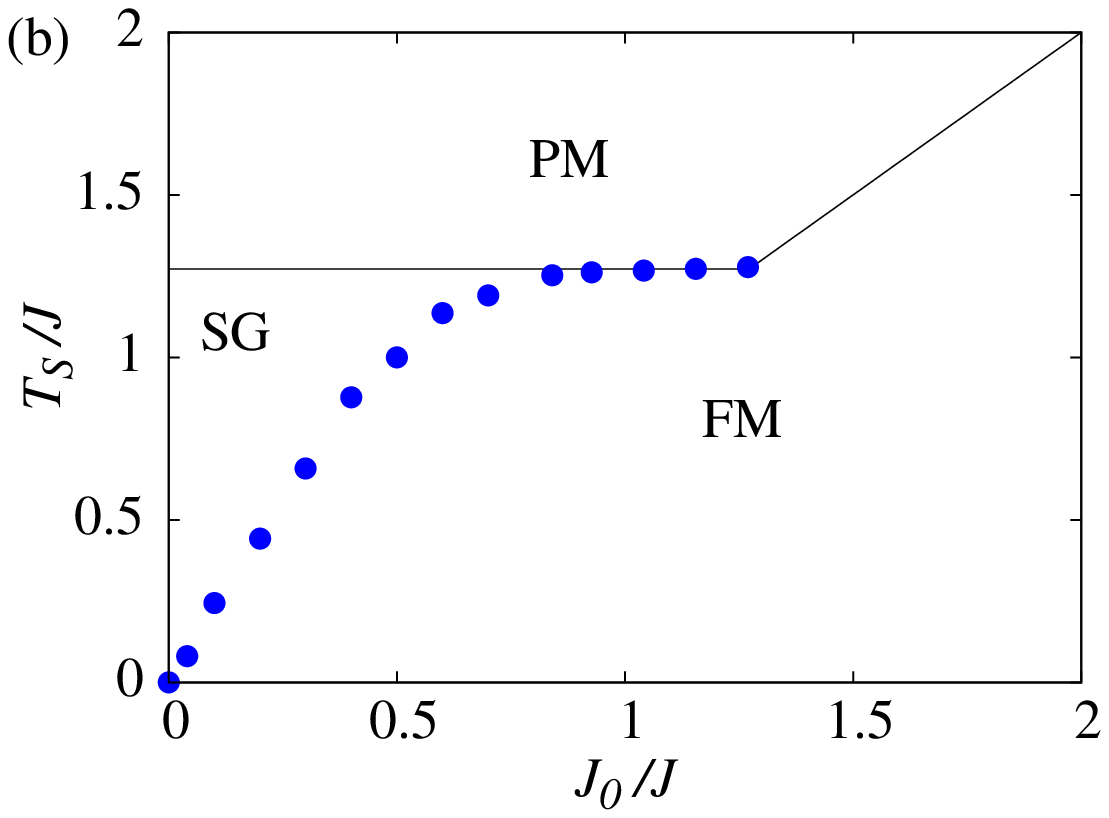}
\end{center}
\caption{(Color online) Phase diagram of a coupled mean-field model for 
(a) $r=1\slash 3$ and (b) $r=3$ with the ratio of two temperatures being $n=2$.
The notations PM, FM and SG mean the paramagnetic, ferromagnetic and
 spin-glass phases, respectively. 
The points $\bullet$ are 
the transition temperature $T_S^{\mbox{\tiny FM-SG}}$  obtained 
by solving the saddle-point equations numerically.
}
\label{n2.0}
\end{figure}
\begin{figure}
\begin{center}
\includegraphics[width=\figwidth]{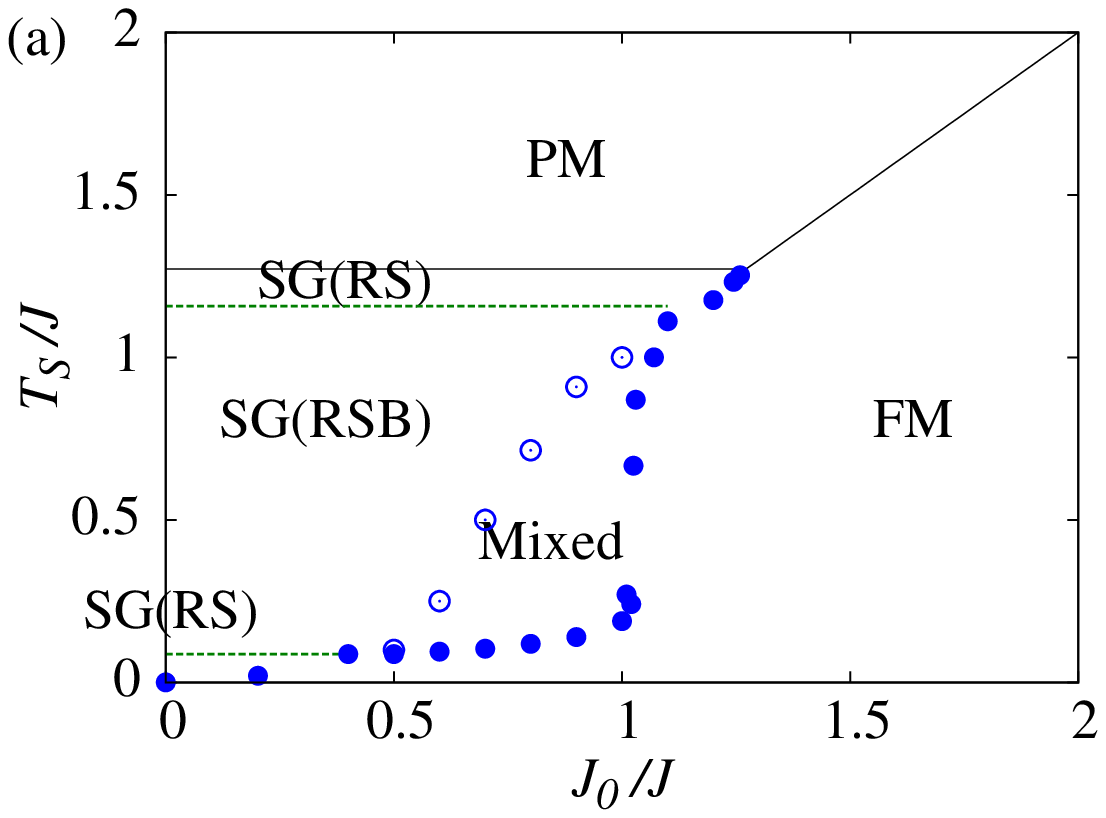}
\includegraphics[width=\figwidth]{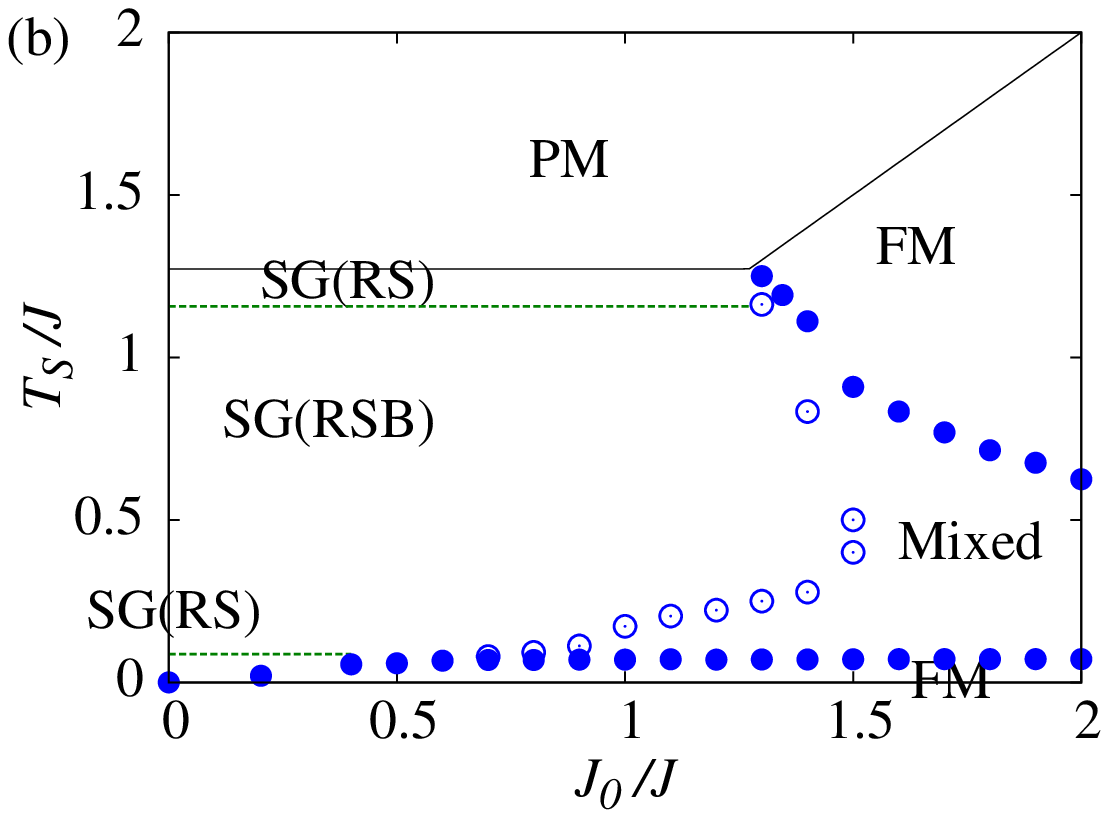}
\end{center}
\caption{(Color online) Phase diagram of a coupled mean-field model with
 $n=0.1$ for (a) $r=1/3$ and (b) $r=3$. 
The open and close circles represent the phase boundary between
 replica-symmetry broken (RSB) SG and 
 mixed phases, and FM and mixed phases, respectively. 
The dashed line is a boundary between replica symmetric (RS) SG and
RSB SG phases. 
}
\label{n0.1}
\end{figure}

As shown in the previous subsection,
the transition temperature $T_S^{\mbox{\tiny FM-SG}}$ between the
ferromagnetic and spin-glass phases depends on the characteristic
parameters $r$ and $n$ of our model.   
Here we carefully discuss $r$ and $n$-dependence of the phase
boundaries. 
To completely obtain the phase boundary of $T_S^{\mbox{\tiny FM-SG}}$, 
we should numerically solve the self-consistent equations
\Eref{saddle_q} and \Eref{saddle_m}.
Some limit cases around multicritical point and near $T_S=0$
can be argued by an expansion of the order parameters. 

At sufficiently low $T_S$, the spin-glass order parameters $q_p$ for
$p=1$ and 2 behave as
\begin{align}
q_p(n)\simeq \left\{
\begin{array}{ll}
1-4\exp(-2\beta_S^2J^2c_p(n-1)),  & \mbox{for}~n> 1\\
\displaystyle 1-\frac{\exp(-\beta_S^2J^2n^2c_p\slash2)}{(1-n)\beta_SJ\sqrt{c_p}} & \mbox{for}~n<1,
\end{array}
\right.
\label{low-temp-exp}
\end{align}
where $c_1=r$ and $c_2=1+r^{-1}$.
They decrease from 1 exponentially in $\beta_S$ for $n>1$ and linearly
in $T_S$ for $n<1$. 
Substituting them into \Eref{FM-SG},
we find that $T_S^{\mbox{\tiny FM-SG}}=nJ_0$ around $J_0=T_S=0$,
irrespective of $r$,  and that the RS solution at $T_S=0$ always
satisfies the stability condition (\ref{AT}).  
Therefore, at $T_S=0$ the ferromagnetic phase stably exists for any
$J_0$ and $n>0$ and it vanishes at $n=0$. 
The latter is recovered to the quench limit studied in
\cite{Takayama-model}.  
At sufficiently low $T_S$, the partial annealing effect yields
the ferromagnetic ordering even for weak $J_0$ and for small ratio $r$
by appropriately selecting $\bm{J}_2$ and $\bm{J}_3$. 
The stability of the ferromagnetic phase near $T_S=0$ is a particular
feature of the partial annealing system of the coupled mean-field
model.  

Near the multicritical point, the spin-glass order parameters can be expressed as
\begin{align}
q_1(n)&\simeq \left\{
\begin{array}{ll}
\frac{2(2\alpha-1)}{(2-n)(\alpha-1)(\alpha c_2-1)}(T_S-\sqrt{\alpha})
 & \mbox{for}~n<2,  \\
\sqrt{\frac{6r^2(2\alpha-1)}{r^2(\alpha-1)+\alpha^3}}(T_S-\sqrt{\alpha})^{1\slash
 2} & \mbox{for}~n=2, 
\end{array}
\right.\\
q_2(n)&\simeq \frac{q_1(n)\alpha}{r}.
\label{mct-exp}
\end{align}
The critical exponent of the spin-glass order parameter is 1 at $n<2$
and $1\slash 2$ at $n=2$. For $n>2$, the order parameter is difficult to
obtain by an expansion because the transition is of first order. 
The phase boundary around the multicritical point significantly depends
on $r$, in contrast to that around $T_S=0$.

The phase diagrams  at $n=2$ for $r=1\slash 3$ and $r=3$ are shown in
\Fref{n2.0} as an example for $n>1$.  
The obtained phase diagram weakly depends on $r$; 
for  $J_0\slash J>\sqrt{\alpha}$, as $T_S$ decreases the transition from the paramagnetic to ferromagnetic phases 
occurs at $T_S^{\mbox{\tiny PM-FM}}$, and for $J_0\slash J<\sqrt{\alpha}$, the
spin-glass phase appears at  $T_S^{\mbox{\tiny
FM-SG}}<T_S<T_S^{\mbox{\tiny PM-SG}}$. 
All the phases found for $n>1$ fulfill the stability condition (\ref{AT}). In particular, the obtained spin-glass phase is correctly
described by the RS solution. It
would be interesting to see that the region of the spin-glass phase for
$r=3$ is reduced as 
compared with that for $r=1\slash 3$. More concretely, 
the transition temperature $T_S^{\mbox{\tiny FM-SG}}$ around the
multicritical point for $r=1\slash 3$ is lower than that for $r=3$.  
This is a counter-intuitive result because
in the case with $r=3$ the majority spins in the system 2 connect with
each other through the spin-glass interactions while the majority spins
are ferromagnetically coupled  in the system 1 in the case with $r=1\slash 3$.  
We shall discuss this point later. 

In addition to the phases found for $n>1$,  other phases appear for
$n<1$, that are the mixed phase characterized by the AT instability with
$m_1>0$ and spin-glass phase with replica symmetry broken(RSB). 
The phase diagram at $n=0.1$ as a typical example for $n<1$ is shown
for $r=1/3$ and $r=3$ in \Fref{n0.1}. 
The mixed phase is found between the ferromagnetic and
spin-glass phases and the region of the mixed phase is enlarged with
increasing $r$. The region of the spin-glass phase is also enlarged with
$r$ and a broken replica symmetric phase exists in the intermediate
$T_S$ region in the spin-glass phase, 
in contrast to
that observed for $n=2$. 
For $r=3$, 
a reentrant transition (PM$\to$FM$\to$Mixed) occurs at $J_0\slash J>\sqrt{\alpha}$.  
Furthermore, the transition from the spin-glass or mixed phases to ferromagnetic phase 
occurs at lower $T_S$, because the ferromagnetic phase is always stable at
$T_S=0$ in our model for any finite $n$. 
Hence,  we can see three successive transitions, PM$\to$FM$\to$Mixed$\to$FM for
$J_0\slash J>\sqrt{\alpha}$. 

\subsection{Reentrant transition}
While  a characteristic feature of our model is found near $T_S=0$,  
another feature is there around the multi-critical point in the phase
diagram. As seen in \Fref{n0.1}(b), a reentrant transition from the
ferromagnetic to a mixed and spin-glass phases occurs near the
multicritical point for $J_0\slash J>\sqrt{\alpha}$. 
This means that the embedded ferromagnetic ordering of the system 1
described by the RS solution becomes unstable as $T_S$ decreases. Such a
reentrant transition has been already reported for the quenched system
with $n=0$ \cite{Takayama-model}.  In this work, we study a partial annealing
effect, namely with finite $n$,  on the stability of the embedded
ferromagnetic ordering. 
The gradient of $T_S^{\rm FM-SG}$ at the multicritical point is regarded
as an indicator of the ``reentrant transition''. Namely, the negative
slope of $T_S^{\rm FM-SG}$ at the multicritical point implies the
existence of the reentrant transition,  although the existence of a
reentrant transition at temperature lower than the multicritical point
is not completely ruled out even in the case with positive slope of
$T_S^{\rm FM-SG}$.   

\begin{figure}
\begin{center}
\includegraphics[width=\figwidth]{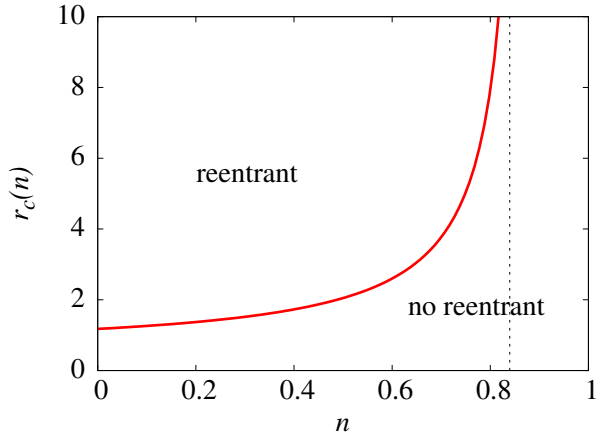}
\end{center}
\caption{
$n$ dependence of the critical ratio $r_c$ above which the reentrant
transition can occurs. 
The ratio $r_c$ diverges at $n_c\simeq 0.8396\cdots$, 
which is shown by the broken line,  with increasing $n$. 
}
\label{reentrant}
\end{figure}

The reentrant transition can occur when the condition
\begin{align}
\frac{{\rm d}J_0^c(T_S)}{{\rm d}T_S}\Big|_{\mbox{\tiny MCP}}<0
\label{grad}
\end{align}
is satisfied where $J^c_0(T_S)=T_S\slash (1+(n-1)q_1(T_S))$ 
and ``MCP'' means the multicritical point. 
Note that the order parameter $q_1$ is a function of $T_S$. 
The RS solution for $q_1$ is sufficient for evaluating the boundary,
because the RS solution is always stable at the multicritical point. 
From \Eref{grad}, the region of $r$ where the reentrant
transition occurs is derived  as
\begin{align}
r>r_c(n),~~~&\mbox{for}~n<n_c\simeq 0.8396\dots 
\end{align}
where the critical ratio $r_c(n)$ is given by 
\begin{align}
r_c(n)=\frac{(2-n)(1+\alpha)}{2(1-n)(2\alpha-1)-(2-n)(\alpha-1)}.
\end{align}
For $n>n_c$, the reentrant transition occurs at $r<r_c(n)$  with
$r_c(n)$ being a negative value, hence it is unphysical. 
The derivation of $r_c(n)$ is based on the assumption
that the transition is of the second order.  
Certainly the transition is of second order for $n\leq 2$, which is 
larger than $n_c$. 
Therefore, the reentrant transition occurs  only for $n<n_c$.
We show $n$-dependence of $r_c(n)$ in \Fref{reentrant}.
At the smaller $n$, the reentrant transition occurs at smaller $r$. 
Eventually, $r_c$ takes $1.618\cdots$ in the quench limit $n=0$,
that is consistent with the results in Ref.~[\onlinecite{Takayama-model}].
Furthermore, it is found that the gradient ${\rm d}J_0^c(T_S)\slash{\rm d} T_S$ at the
multicritical point is a monotonically increasing function of $r$ for $n>n_c$,
and this fact yields the shrink of the spin-glass phase as $r$
increases, as shown in \Fref{n2.0}.
Thus,  the stability of the embedded ferromagnetic ordering of the
system 1 depends quantitatively on the value of $n$ that is a relative
temperature $\beta_J/\beta_S$.

Here, we argue a typical feature of the partially annealed interaction
$\bm{J}_3$ from the obtained phase diagram and the critical ratio
$r_c(n)$. In equilibrium of the partial annealing system, the
interactions $J_{ij}$ are in general proportional to $n\langle
S_iS_j\rangle$ given as a most probable value in $P(\bm{J})$. 
Supposing that the order parameter $q_2$ of the system 2 has a finite
value at sufficiently low temperature $T_S$,  
the spins in the system 1 are subjected to an effective field $r\sqrt{q_2}$ caused
by the system 2 through the interactions $\bm{J}_3$. 
For $n<n_c$, the interactions $\bm{J}_3$ are considered to be almost
random according to the bare distribution $P(\bm{J}_0)$, and hence an 
effective field is a random field for the system 1. 
This random effective field also does not favor the ferromagnetic order of
the system 1, and the spin glass phase is enhanced as $r$ increases. 
As a consequence of the competition between $r$ and $J_0$, the reentrant
transition at $J_0\slash J>\sqrt{\alpha}$ is found at $r_c$. 
For $n>n_c$, the interactions $\bm{J}_3$ likely have the same sign as
$\langle S_{1,i}S_{2,j}\rangle$. Then, the effective field from the
system 2 to the system 1 is not random but it supports the ferromagnetic
order in the system 1. 
Therefore, the ferromagnetic ordering is stabilized even as $r$
increases and the system 2 with the spin-glass couplings dominates. 
The reentrant transition does not occur for any $r$ for $n>n_c$.

\subsection{local structures of frustration}
\label{simulation}

In this subsection, we discuss equilibrium properties of the coupled
mean-field model from the view point of interactions $\bm{J}$. 
Frustration is a key quantity that characterizes a structure of the
interactions in spin glasses and related random spin models. 
It is defined as a product of $J_{ij}$s 
along a minimal loop, whose length is three in the fully-connected model
studied in this work. 
If the interactions among three spins satisfy the
condition $J_{ij}J_{jk}J_{ki}<0$, three terms of the local energy  cannot be minimized
simultaneously.
Such interactions are called to have frustration \cite{Toulouse}.
Meanwhile, all the interactions satisfying $J_{ij}J_{jk}J_{ki}>0$
do not have frustration and the energy of the spins attains
the global minimum value although relative directions of spins are not
align totally. This type of interactions is called Mattis model \cite{Mattis} and
all of the interaction sets can be reduced to the pure ferromagnetic
model by local gauge transformation \cite{Nishimori}.

In the coupled mean-field model, we should consider three distinct
frustration parameters originated from three types of interactions, as
shown in \Fref{takayama-frust}, are defined; 
\begin{align}
\Phi_{(2)}&=\sum_{i<j<k}J_{2,ij}J_{2,jk}J_{2,ki}, \\
\Phi_{(3a)}&=\sum_{i<j,k}J_{1,ij}J_{3,jk}J_{3,ki}, \\
\Phi_{(3b)}&=\sum_{i,j<k}J_{3,ij}J_{2,jk}J_{3,ki}.
\end{align}
In equilibrium, these parameters are to be taken average over the
equilibrium distribution $P(\bm{J})$ in \Eref{eqn:eqJ} as
$[\Phi_{(2)}]_n$, $[\Phi_{(3a)}]_n$ and $[\Phi_{(3b)}]_n$ with $[\cdots
]_n$ being the average with respect to $\bm{J}$. For the limit
$n\rightarrow 0$ with $\beta_S$ kept finite, the distribution
$P(\bm{J})$ is identical to the bare distribution $P_0(\bm{J})$ and the
average $[\cdots ]_n$ is reduced to $[\cdots ]_0$. In this limit, the
frustration parameters become zero. 
If the frustration parameters take a positive finite value at finite
$n$, this indicates that the frustration is decreased as a consequence of
correlation of $\bm{J}$. 

When the bare distribution of the interactions is Gaussian, 
the averaged frustration parameters are expressed in terms of the order parameters
\cite{Dotsenko,SH-unpublished}.
In this model, the averaged frustration parameters under the RS ansatz are described as follows,
\begin{align}
[\Phi_{(2)}]_n&=\beta_S^3\{\lambda_2^3+(n-1)\mu_2^3\}, \label{eq:2}\\
[\Phi_{(3a)}]_n&=\beta_S^2nJ_0\lambda_2m_1^2, \label{eq:3a}\\
[\Phi_{(3b)}]_n&=\beta_S^3\{\lambda_1\lambda_2^2+(n-1)\mu_1\mu_2^2\},
 \label{eq:3b}
\end{align}
where $\lambda_p=1+(n-1)q_p$ and $\mu_p=1-q_p$ are the eigenvalues of the
$n\times n$ matrix $Q_p$ whose diagonal components are 1 and off-diagonal components are $q_p$.
We can also derive the frustration parameter with multiple step RSB,
that does not yield a significant quantitative change \cite{SH-unpublished}.

At finite $n$, $[\Phi_{(2)}]_n$  and $[\Phi_{(3b)}]_n$ take a finite
value depending on $\beta_S$, while $[\Phi_{(3a)}]_n$ zero in the
paramagnetic phase. This moderate decrease of the frustration is considered
to be due to an emergence of ``short-range'' correlation of $\bm{J}$
induced by the partial annealing. A considerably qualitative change of
the frustration parameters is accompanied by phase transitions. 
In the spin-glass phase with $q_1>0$, $q_2>0$ and $m_1=0$,
the frustration parameters $[\Phi_{(2)}]_n$ and $[\Phi_{(3b)}]_n$ largely
increase, but $[\Phi_{(3a)}]_n$ is still zero.
Therefore, the frustration is non-uniformly distributed in the system and
a ``local'' structure of frustration is formed;
the frustrations in the system 2 and a part of $\bm{J}_3$ decrease,
but a remaining part of $\bm{J}_3$ has frustration as much as the
randomly constituted interactions. 
The local structure prefers to decrease selectively the frustration of
the system 2 in the spin-glass phase, and does not cooperative with the ferromagnetic state of
the system 1.
Meanwhile, all of the frustration parameters take a positive value in the ferromagnetic phase, 
because in this phase all order parameters are finite.
In this case, all interactions $\bm{J}_2$ and $\bm{J}_3$
decrease the energy of the ferromagnetic state of the system 1, 
meaning that  the effective field from the system 2 energetically
supports the ferromagnetic state of the system 1. 
This type of interactions is similar to that of the Mattis states \cite{Mattis}.
These observations of the frustration parameters certify the
validity of 
the argument of $r_c(n)$ and $n_c$ in the previous section.

Before closing this section, we discuss the order of the frustration parameters.
The interactions are of order of $N^{-1\slash 2}$ in the bare
distribution.  When the frustrations completely vanish with keeping the order of $J_{ij}$, the order of the 
frustration parameters become $O(N^{3\slash 2})$ by definition. 
However, the averaged frustration parameters should be $O(1)$ quantities for
any $T_S$, as seen from the expressions \Eref{eq:2}, \Eref{eq:3a} and
\Eref{eq:3b}. This means that the order of $J_{ij}$s is appropriately
modified to $O(N^{-1})$ through the partial annealing, and
the extensivity of thermodynamic quantities such as energy and
free energy is maintained at whole temperature region.
It is a characteristic of the partial annealing with the bare
distribution function of $\bm{J}$ being Gaussian.  
When the bare distribution is bimodal and the allowed value of
$\bm{J}$ is restricted to $\pm J$, namely $\pm\bm{J}$ model,
the order of $J_{ij}$ cannot be changed and the energy becomes
$O(N^{3\slash 2})$ in the resultant ferromagnetic phase,
that is the lack of extensivity. We should be careful for the study
of thermodynamics in such models.

\section{Summary and discussion}
\label{conclusion}

We have studied equilibrium properties of the coupled mean-field model
in the partial annealing system, in which the system 1 with an embedded
ferromagnetic ordering and the system 2 with no embedded pattern are
coupled with spin-glass interactions. 
In this model, the interactions  as well as spins are regarded as dynamical variables,
but their time-scales are completely separated from each other. 
The spins $\bm{S}$ and interactions $\bm{J}$ touch to their own heat
bath with different temperatures $T_S$ and $T_J$. By using the replica
method, the free energy of the system is derived as functions of the two
temperatures,  and 
the phase diagram is obtained in the two-temperature plane.  
The phase boundary between the ferromagnetic
phase and the spin-glass phase interestingly depends on the model parameters,
the size ratio $r$ of the system 2 to the system 1 and the ratio $n$ between two temperatures.  

We carefully studied $T_S^{\mbox{\tiny FM-SG}}$ around the multicritical
point. It is found that there exists a critical value $n_c$ in the ratio
$n$, which characterizes if the reentrant transition occurs around
the multicritical point. For $n<n_c$, the partial annealing effect on
the inter-coupling $\bm{J}_3$ 
is not significant to enhance the embedded ferromagnetic ordering in the
system 1. The effective field  from the system 2 to  the system 1 does not
support the ferromagnetic ordering in the system 1. 
Therefore for a fixed $n<n_c$ as the ratio $r$ increases, 
in other words the number of spins in the
system 2 increases relatively,
a competition about the ferromagnetic order is expected;
the random filed destabilizes the ferromagnetic order and the coupling
$J_0$ stabilizes it. As a result of the competition, the ferromagnetic
order is destabilized eventually. 
This can be seen as the reentrant transition in
the phase diagram. 
Meanwhile, for $n>n_c$ the effective field 
supports to achieve the ferromagnetic ordering in the system 1 irrespective of the value of $r$, and
the spin-glass region is narrowed monotonically as $r$ increases.


We introduced three frustration parameters, 
$\Phi_{(2)}$,  $\Phi_{(3a)}$ and $\Phi_{(3b)}$ for characterizing
partially annealed interactions of the coupled mean-field model.  
They are expressed in terms of the order parameters 
of the spin system in equilibrium.  
Using the parameters, we classify the type of the interactions in this
model into three categories,  that correspond to each phase;
interactions with a local structure of frustration in the spin-glass phase,
Mattis-like interactions in the ferromagnetic phase, and
randomly constructed interactions in the paramagnetic phase. 
The characteristic interactions found in the spin-glass phase do not
cooperatively support the ferromagnetic ordering of the system 1, 
that is to say,
the frustration is eliminated from the system 2 independently of the
ferromagnetic interactions in the system 1. 

This is quite different from those obtained in a related model
previously studied \cite{SHK,SHK-PRE}, in which a locally embedded
pattern is supported  by the interactions surrounding the pattern 
and the frustration in the rest of the interactions still remains 
extensively. 
Although both models have the same structure
that an ordering pattern is embedded in a part of the system,
the partial annealing leads to the different property of interactions;
the interactions in the previously studied model support the embedded pattern,
and those in this study disturb it.
A reason of the difference may be originated from the different
choice of $H_J$.  In the present work, $H_J(\bm{J})$ contains the spin
free energy, while $H_J(\bm{J})$ in \cite{SHK,SHK-PRE} has a local
order parameter. 
The results suggest that 
an explicit form of $H_J(\bm{J})$ strongly affects on the construction
of the interaction $\bm{J}$ and the ordering of the spin $\bm{S}$, in
particular, an entropic effect of $\bm{J}$ is nontrivial. 
For proper understanding of the constructed interactions in the partial
annealing, we have to pursue some variant models with  a general type of
$H_J$. However, analytical studies of the partial annealing
to this time heavily
relies on the replica method, that is applicable to only systems with the
spin free energy as $H_J$. 
The partially annealed system with $H_J$ that does not contain the spin
free energy has yet to be revealed in terms of spin-glass theory.
An extended formalism of the partially annealed system is required for
further applications to biological or engineering models.  

In this work, we focus our attention to the mean-field analysis based on
the fully connected spin models. When we introduce a diluted model such
as the Viana-Bray model \cite{VianaBray} in partial annealing, the geometric structure of the partially
annealed $\bm{J}$ may play an important role in cooperative phenomenon. 
Furthermore, a decoupling transition of two degrees of freedom might
occur in the diluted models, while the transition of frustration is completely
correlated to the transition of the spin variables in our model. 
Some work in this direction is in progress. 


We end with an account of a perspective of the partially annealed system. 
In most of the studies concerned with the partial annealing, the free
energy is used for Hamiltonian $H_J(\bm{J})$ of the slow variables
$\bm{J}$. This yields a replicated system with a finite replica
number determined by the ratio of two temperatures in the partially
annealed system.  The partial annealing can be considered to give a
physical meaning to the replica approach before taking a limit of the
replica number to zero \cite{PCS}. Meanwhile, the large
deviations of the free energy of mean-field spin glasses are 
studied by the replica method with a finite replica
number \cite{Parisi-Rizzo}.  
A mechanism of replica symmetry breaking is studied through a phase
transition as the replica number is taken to zero \cite{TNakajima}. 
Thus, the study of finitely-replicated system can provide new insights
into the spin-glass theory. 
In this work, we discussed a phase transition as a cooperative phenomena
of the fast and slow variables with the replica number varying, that might
give a different viewpoint to the finitely replicated system. 
It would be interesting to classify possible universality classes of
phase transitions of the finitely replicated systems, particularly  in
finite dimensions.



\begin{acknowledgments}
We would like to thank C.~H.~Nakajima and T.~Nakajima for helpful comments and discussions.
This work was supported by a Grant-in-Aid for Scientific Research (No.18079004) from MEXT
and JSPS Fellows (No.20$-$10778) from JSPS. 
\end{acknowledgments}

\newpage 

\end{document}